\documentclass[twocolumn]{aastex62}
\usepackage{graphicx}

\newcommand{\Mjup}{\mbox{$\rm M_\mathrm{Jup}$}}
\newcommand{\Msun}{\mbox{$\rm M_{\odot}$} }
\newcommand{\Rsun}{\mbox{$\rm R_{\odot}$} }
\newcommand{\Lsun}{\mbox{$\rm L_{\odot}$} }
\newcommand{\e}[1]{\times 10^{#1}}
\shorttitle{Resolved Millimeter Observations of the HD 170773 Debris Disk}
\shortauthors{Sepulveda et al.}

\begin{document}
\title{The REASONS Survey: Resolved Millimeter Observations of a Large Debris Disk Around the Nearby F Star HD 170773}
\correspondingauthor{Aldo G. Sepulveda}
\email{aldogsepulveda@gmail.com}
\author[0000-0002-8621-2682]{Aldo G. Sepulveda}
\affiliation{Department of Physics \& Astronomy, The University of Texas at San Antonio
1 UTSA Circle
San Antonio, TX 78249, USA}
\author[0000-0003-4705-3188]{Luca Matr\`a}
\altaffiliation{Submillimeter Array (SMA) Fellow}
\affiliation{Harvard-Smithsonian Center for Astrophysics 
60 Garden St. 
Cambridge, MA 02138, USA}
\author{Grant M. Kennedy}
\affiliation{Department of Physics, University of Warwick, Gibbet Hill Road, Coventry, CV4 7AL, UK}
\affiliation{Centre for Exoplanets and Habitability, University of Warwick, Gibbet Hill Road, Coventry, CV4 7AL, UK}
\author{Carlos del Burgo}
\affiliation{Instituto Nacional de Astrof\'\i sica, \'Optica y Electr\'onica, Luis Enrique Erro 1, Sta. Ma. Tonantzintla, 72840 Puebla, Mexico}
\author{Karin I. \"Oberg}
\affiliation{Harvard-Smithsonian Center for Astrophysics 
60 Garden St. 
Cambridge, MA 02138, USA}
\author{David J. Wilner}
\affiliation{Harvard-Smithsonian Center for Astrophysics 
60 Garden St. 
Cambridge, MA 02138, USA}
\author{Sebasti\'an Marino}
\affiliation{Max Planck Institute for Astronomy, K\"onigstuhl 17, 69117 Heidelberg, Germany}
\author[0000-0001-8568-6336]{Mark Booth}
\affiliation{Astrophysikalisches Institut und Universit\"atssternwarte, Friedrich-Schiller-Universit\"at Jena, Schillerg\"a\ss chen 2-3, 07745 Jena, Germany}
\author[0000-0003-2251-0602]{John M. Carpenter}
\affiliation{Joint ALMA Observatory (JAO), Avenida Alonso de C\'ordova 3107 Vitacura 7630355, Santiago, Chile}
\author[0000-0001-9764-2357]{Claire L. Davies}
\affiliation{Astrophysics Group, School of Physics, University of Exeter, Exeter EX4 4QL, UK}
\author{William R.F. Dent}
\affiliation{Joint ALMA Observatory (JAO), Avenida Alonso de C\'ordova 3107 Vitacura 7630355, Santiago, Chile}
\affiliation{European Southern Observatory, Avenida Alonso de C\'ordova 3107 Vitacura 7630355, Santiago, Chile}
\author{Steve Ertel}
\affiliation{European Southern Observatory, Avenida Alonso de C\'ordova 3107 Vitacura 7630355, Santiago, Chile}
\affiliation{Large Binocular Telescope Observatory, 933 North Cherry Avenue, Tucson, AZ 85721, USA}
\affiliation{Steward Observatory, Department of Astronomy, University of Arizona, 993 N. Cherry Ave, Tucson AZ, 85721, USA}
\author{Jean-Francois Lestrade}
\affiliation{LERMA, Observatoire de Paris, PSL, CNRS, Sorbonne Universit\'es, UPMC Univ., 61 avenue de l'Observatoire, Paris, France}
\author[0000-0001-6208-1801]{Jonathan P. Marshall}
\affiliation{Academia Sinica, Institute of Astronomy and Astrophysics, 11F Astronomy-Mathematics Building, NTU/AS campus, No. 1, Section 4, Roosevelt Rd., Taipei 10617, Taiwan}
\author{Julien Milli}
\affiliation{European Southern Observatory, Avenida Alonso de C\'ordova 3107 Vitacura 7630355, Santiago, Chile}
\author{Mark C. Wyatt}
\affiliation{Institute of Astronomy, University of Cambridge, Madingley Road, Cambridge CB3 0HA, UK}
\author{Meredith A. MacGregor}
\altaffiliation{NSF Astronomy and Astrophysics Postdoctoral Fellow}
\affiliation{Department of Terrestrial Magnetism, Carnegie Institution for Science, 5241 Broad Branch Road, Washington, DC 20015, USA}
\author{Brenda C. Matthews}
\affiliation{Department of Physics \& Astronomy, University of Victoria, 3800 Finnerty Road, Victoria, BC, V8P 5C2, Canada}
\affiliation{Herzberg Astronomy \& Astrophysics Programs, National Research Council of
Canada, 5071 West Saanich Road, Victoria, BC, V9E 2E7, Canada}

\begin{abstract}
Debris disks are extrasolar analogs to our own Kuiper Belt and they are detected around at least 17\% of nearby Sun-like stars. The morphology and dynamics of a disk encode information about its history, as well as that of any exoplanets within the system. We used ALMA to obtain 1.3 mm observations of the debris disk around the nearby F5V star HD 170773. We image the face-on ring and determine its fundamental parameters by forward-modeling the interferometric visibilities through a Markov Chain Monte Carlo approach. Using a symmetric Gaussian surface density profile, we find a 71 $\pm$ 4 au wide belt with a radius of 193$^{+2}_{-3}$ au, a relatively large radius compared to most other millimeter-resolved belts around late A / early F type stars. This makes HD~170773 part of a group of four disks around A and F stars with radii larger than expected from the recently reported planetesimal belt radius - stellar luminosity relation. Two of these systems are known to host directly imaged giant planets, which may point to a connection between large belts and the presence of long-period giant planets. We also set upper limits on the presence of CO and CN gas in the system, which imply that the exocomets that constitute this belt have CO and HCN ice mass fractions of $<77$\% and $<3$\%, respectively, consistent with Solar System comets and other exocometary belts.
\end{abstract}

\keywords{circumstellar matter --- planet-disk interactions --- stars: individual (HD 170773) --- techniques: interferometric}

\frenchspacing
\section{Introduction}\label{sec:intro}
Debris disks (also known as exocometary belts or planetesimal belts) are rings of dust, exocomets, and planetesimals analogous to our Solar System's Kuiper Belt and are detected around at least $\sim$17\% of Sun-like stars \citep{Montesinos2016,Sibthorpe18}. Debris disks span different regions across planetary systems, including outer regions (colder belts typically located at tens of au from their host star) to inner regions (warmer belts at a few au that are generally more difficult to detect) \citep{kennedy&wyatt2014,ballering2017}. These debris disks are intimately related to the formation of exoplanets and offer clues about the evolution and dynamical history of the system \citep[][]{bowler2016,Wyatt2018haex.book}. Debris disks are generally thought to be maintained by a collisional cascade process, where destructive collisions between larger planetesimal bodies produce smaller planetesimals that further collide to produce the small dust grains that are observed \citep[e.g.][and references therein]{wyatt2011springer, hughes18}. 

The constituent dust grains reprocess starlight into radiation at longer wavelengths that are comparable to the sizes of the emitting bodies. The smallest dust grains are continuously blown out of the disk by the radiation pressure of the host star and additionally by stellar winds \citep{backman&paresce1993}.
Thus, imaging debris disks in the infrared highlights emission (and scattering) from small grains that are blown out and may not delineate the true spatial architecture of the parent planetesimal belts. To probe the location of larger dust and planetesimals too massive to be strongly influenced by stellar winds and radiation pressure, observations must be made at longer wavelengths in the millimeter/submillimeter regime and must spatially resolve the disk \citep[e.g.][]{augereau2001}.

Understanding why debris disks form at their particular radii from their host stars is key to better understanding the physical mechanisms that create these disks. \citet{matra2018b} conducted a population study of 26 millimeter-resolved debris disks and found a statistically significant correlation between the host star luminosity and disk radius, which persists when accounting for potential observational biases. Constraining the radii for additional debris disks is the next step to further characterize this radius-luminosity relationship and to further explore other potential correlations. Empirically quantifying any correlations between disk parameters and host star properties is an essential step to test and refine planet and debris disk formation models. The REASONS (REsolved ALMA and SMA Observations of Nearby Stars) survey, the follow up of the JCMT SCUBA-2 Observations of Nearby Stars (SONS) Legacy survey \citep{holland2017}, aims to approximately double the sample size of millimeter-resolved debris disks.

We present new 1.3 mm observations of the debris disk around the nearby \citep[37.02 $\pm$ 0.06 pc:][]{Bailer-Jones2018,gaia, gaiaDR2_2018} F5V \citep{Gray06} star HD 170773. The infrared excess of HD 170773 was first detected by \citet{sadakane86} with \textit{IRAS}, which is a general indication that a debris disk might be present in the system.  \citet{Nilssonetal2010} first detected the disk at submillimeter wavelengths (870$\mu$m) using the APEX telescope and derived a disk radius of 170 au from the flux distribution. The disk was also detected and resolved at various far-IR wavelengths (70$\mu$m, 100$\mu$m, 160$\mu$m) by \citet{mooretal2015} using \textit{Herschel Space Observatory}, with the weighted average disk radius reported to be 173.4 $\pm$ 2.8 au. \citet{holland2017} then resolved the disk at 850$\mu$m using the JCMT and estimated the disk radius as 252 $\pm$ 26 au.
We spatially resolve the debris disk around HD 170773 in the millimeter regime to probe the spatial properties of the parent belt that is traced by the millimeter dust grains too massive to be influenced by the stellar radiation and activity. 

In \S\ref{sec:obs} we describe our ALMA observations of HD 170773. \S\ref{sec:res} details our analysis of the observations and the process of modeling the disk and constraining the fundamental parameters in a Bayesian fashion using Markov Chain Monte Carlo (MCMC). Here we check the ALMA observations for CO and CN gas and set upper limits on the corresponding mass fractions of the exocomets. We also infer the stellar parameters of HD~170773 with a Bayesian method and using \textit{Gaia}. We discuss the results in \S\ref{sec:discuss} and describe some of the consequences of our parameter constraints. 

\begin{deluxetable*}{cccc}
\tablehead{
\colhead{Parameter} & \colhead{2018 Apr 29} & \colhead{2018 May 03} & \colhead{2018 Jun 05} 
}
\caption{Summary of ALMA Observation Parameters \label{tab:obs}}
\startdata
No. Antennas&10&12&50\\
Antenna size&7 m&7 m&12 m\\
Time on Target&34m 49s&34m 49s&14m 12s\\
J2000 Pointing R.A.&18h 33m 01.056s&18h 33m 01.056s&18h 33m 01.057s\\
J2000 Pointing Dec.&$-$39$^{\circ}$ 53\arcmin\ 32.744\arcsec&$-$39$^{\circ}$ 53\arcmin\ 32.745\arcsec&$-$39$^{\circ}$ 53\arcmin\ 32.752\arcsec\\
Min/Max baseline&8.9 to 48.9 m&8.9 to 48.9 m&15.0 to 360.6 m\\
Min/Max PWV&1.30 to 1.87 mm&0.42 to 0.63 mm&1.38 to 1.50 mm\\
Gain Calibrator&J1802$-$3940&J1802$-$3940&J1802$-$3940\\
Passband/Flux Calibrator&J1924$-$2914&J1924$-$2914&J1924$-$2914\\
Primary Beam FWHM&46\farcs0&46\farcs0&26\farcs9\\
\enddata
\end{deluxetable*} 
\section{ALMA Observations} \label{sec:obs}

Three ALMA Band 6 observations at 1.3 mm (211 - 275 GHz) of HD 170773 were made on 2018 Apr 29, May 3 and Jun 5. The phase center of the observations is at the proper motion-corrected J2000 stellar position. The interferometric visibility data was fully calibrated by the ALMA observatory using their pipeline.  Table \ref{tab:obs} summarizes the observing parameters. The ALMA interferometer samples the Fourier transform of the sky brightness distribution, resulting in complex valued visibilities in u-v space. Observing HD~170773 with both the Atacama Compact Array (ACA) and the 12-meter array provides greater coverage of the u-v visibility space that in turn better recovers extended emission in the disk image. 

The correlator setup included four 2 GHz-wide spectral windows, two of which were centered at 243.1 and 245.1 GHz at low spectral resolution for continuum, although all four were used in obtaining the HD~170773 continuum image. To obtain the dust continuum image from the visibilities in u-v space, we first average the observations in frequency to reduce the data size, then concatenate all three observations. We then use the \texttt{tclean} task of \texttt{CASA} 5.4.1 \citep{CASA} to inverse transform and deconvolve the visibility data into the 1.3 mm CLEAN continuum emission image. We apply a Gaussian u-v taper of 2\arcsec and a Briggs weighting factor of 0.5 to the visibilities in the imaging process. The Gaussian u-v taper boosts the SNR of the ring at the cost of resolution. 

The other two spectral windows were centered at 230.1 and 227.2 GHz at high spectral resolution to cover the CO J=2-1 transition (at 230.538 GHz) and the CN N=2-1 transitions (where we focus on the strongest fine and hyperfine structure transition, at 226.875 GHz). The spectral resolution of these spectral windows is $\sim$1.28 km/s (for a channel width of 488.281 kHz or 0.64 km/s). We first subtracted continuum emission from the visibilities using the \texttt{CASA} \texttt{uvcontsub} task, then proceeded to imaging (with the same weighting and u-v taper as the continuum dataset) to produce CO and CN datacubes covering $\pm$50 km/s from the expected stellar radial velocity \citep[$-16\pm1$ km/s in the heliocentric frame,][]{gaiaDR2_2018}.

\section{Analysis} \label{sec:res}
\subsection{1.3 mm Dust Continuum Image} \label{subsec:cont}
\begin{figure}
    \includegraphics[trim=1.4cm 0cm 1cm 0cm, clip, width=3.25in]{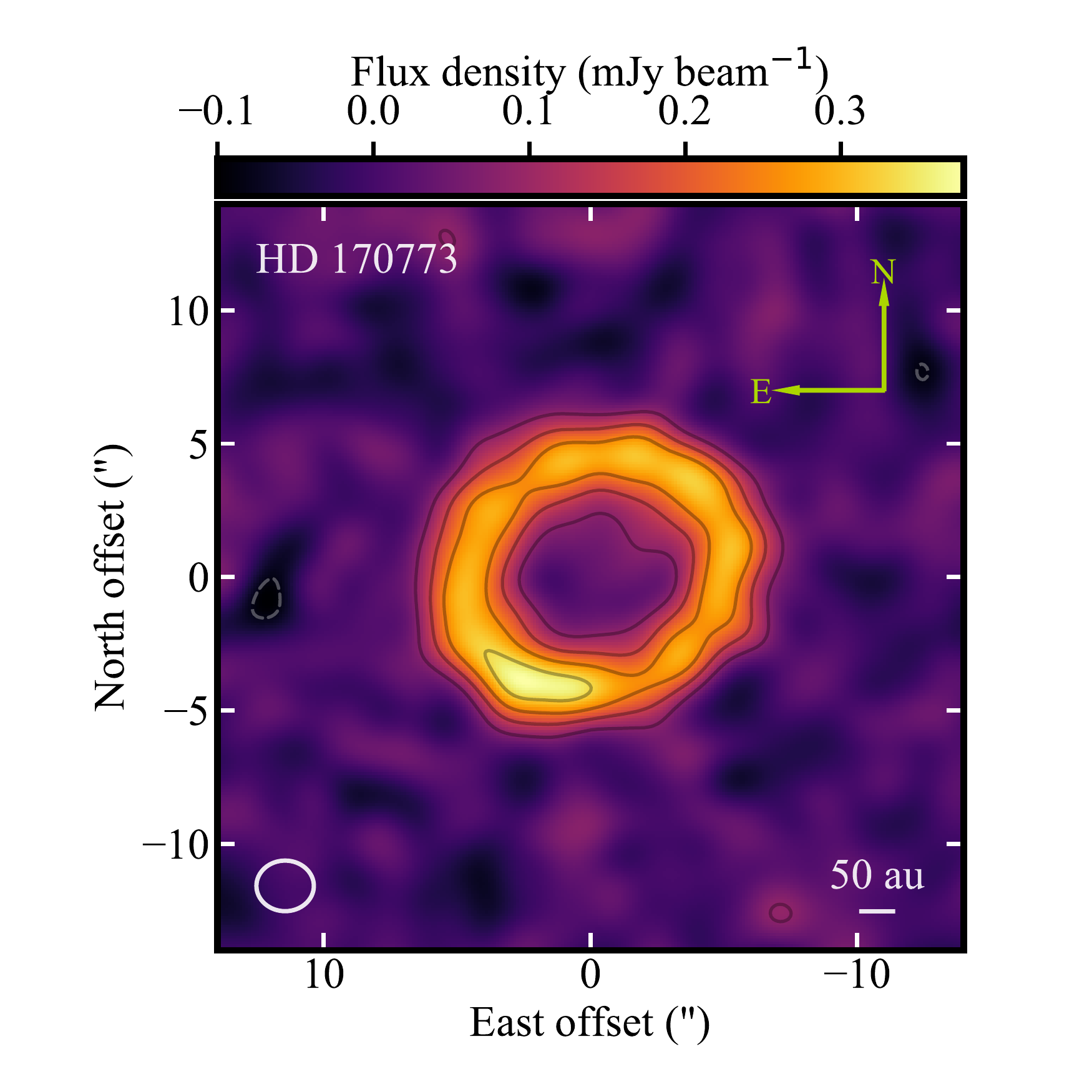}
    \centering
    \caption{Combined ACA + 12-meter array CLEAN image of HD 170773 dust continuum emission at 1.3 mm where a Gaussian u-v taper of 2\arcsec and a Briggs weighting factor of 0.5 are used. Contours are drawn at the levels: $[-3,3,6,9,12] \times (\sigma_{\rm RMS} =$ 2.8$\e{-2}$ mJy beam$^{-1}$). The beam size is indicated in the lower left corner of the image and measures 2\farcs15 by 1\farcs90 with a position angle of 89.9$^{\circ}$ (east of north). 
    }
    \label{fig:emission}
\end{figure}
Figure \ref{fig:emission} shows the 1.3 mm continuum emission of HD~170773. The noise level, measured as the RMS in a region free from emission, is $\sigma_{\rm RMS} =$ 2.8$\e{-2}$ mJy beam$^{-1}$  where the beam size is 2\farcs15 by 1\farcs90. The peak signal is 3.8$\e{-1}$ mJy beam$^{-1}$, yielding a peak SNR of 14. The disk is close to face-on and the distance from the host star to the radial emission peak along the disk major axis is about $\sim$5\arcsec ($\sim$185 au). Although we use a u-v taper to boost the S/N of the image (\S \ref{sec:obs}), the width is clearly resolved when imaging with no u-v taper. The continuum image shows a tentative asymmetry where the SE disk emission may be more pronounced. However, the measured peak signal in the SE quadrant is only $\sim$2$\sigma_{\rm RMS}$ higher than the peak signal measured in the NW quadrant, which means there is not significant evidence for an asymmetry. We also compared the flux density of the NW and SE halves of the disk and found no significant difference. 

\begin{deluxetable}{ccccc}
\tablehead{
\colhead{Instrument/} & \colhead{Waveband} & \colhead{Photometry}& \colhead{Unit} & \colhead{Reference}\\
\colhead{Filter}& \colhead{($\mu$m)}} 
\caption{Photometry used to generate Fig. \ref{fig:SED}.\label{tab:photometry}} 
\startdata
$U-B$\tablenotemark{a}&...&-0.06$\pm$0.03&mag&1\\
Str\"omgren $c_1$\tablenotemark{a}&...&0.48$\pm$0.02  &mag&2\\
B$_T$&0.42&6.71$\pm$0.02&mag&3\\
Str\"omgren $m_1$\tablenotemark{a}&...&0.15$\pm$0.01&mag&2\\
$B-V$\tablenotemark{a}&...&0.42$\pm$0.02&mag&1\\
$b-y$\tablenotemark{a}&...&0.28$\pm$0.01&mag&2\\
V$_T$&0.53&6.27$\pm$0.01&mag&3\\
H$_P$&0.54&6.32$\pm$0.01&mag&4\\
\textit{V}&0.55&6.23$\pm$0.02&mag&1\\
J&1.2&5.42$\pm$0.03&mag&5\\
H&1.6&5.28$\pm$0.04&mag&5\\
K$_S$&2.2&5.20$\pm$0.02&mag&5\\
\textit{WISE} W1&3.4&5.21$\pm$0.14&mag&6\\
\textit{WISE} W2&4.6&5.05$\pm$0.06&mag&6\\
\textit{AKARI}/IRC&9.0&490$\pm$14&mJy&7\\
\textit{WISE} W3&12&5.22$\pm$0.05&mag&6\\
\textit{WISE} W4&22&5.08$\pm$0.07&mag&6 \\
\textit{Spitzer}/MIPS&24&67$\pm$1&mJy&8\\
\textit{Herschel}/PACS&70&794$\pm$24&mJy&9\tablenotemark{b}\\
\textit{Spitzer}/MIPS&70&788$\pm$79&mJy&10\\
\textit{Herschel}/PACS&100&1071$\pm$67&mJy&9\tablenotemark{b}\\
\textit{Herschel}/PACS&160&863$\pm$44&mJy&9\tablenotemark{b}\\
\textit{Herschel}/SPIRE&250&358$\pm$26&mJy&11\\
\textit{Herschel}/SPIRE&350&177$\pm$16&mJy&11\\
JCMT/SCUBA-2&450&$<$135&mJy&12\\
\textit{Herschel}/SPIRE&500&67$\pm$10&mJy&11\\
JCMT/SCUBA-2&850&26$\pm$2&mJy&12\\
APEX/LABOCA&870&18$\pm$5&mJy&13\\
ALMA&1300&6.2$\pm$0.6&mJy&9\\
\enddata
\tablenotetext{a}{Colors (i.e. flux ratios) are fit directly.}
\tablenotetext{b}{PACS fluxes were derived with apertures as described in \citet{Sibthorpe18}.}
\tablereferences{1: \citet{2006yCat21680M}; 2:  \citet{2015AA580A23P}; 3: \citet{2000AA355L27H}; 4: \citet{1997ESASP1200E}; 5: \citet{2003tmcbookC}; 6: \citet{2010AJ1401868W}; 7: \citet{2010AA514A1I}; 8: IRSA, \url{https://irsa.ipac.caltech.edu}; 9: This work; 10: \citet{2014ApJS21125C}; 11: \citet{2017Spire} ; 12: \citet{holland2017}; 13: \citet{Nilssonetal2010}}
\end{deluxetable} 

\begin{figure}
    \includegraphics[trim=0.2cm 0cm 0cm 0cm, clip, width=3.5in]{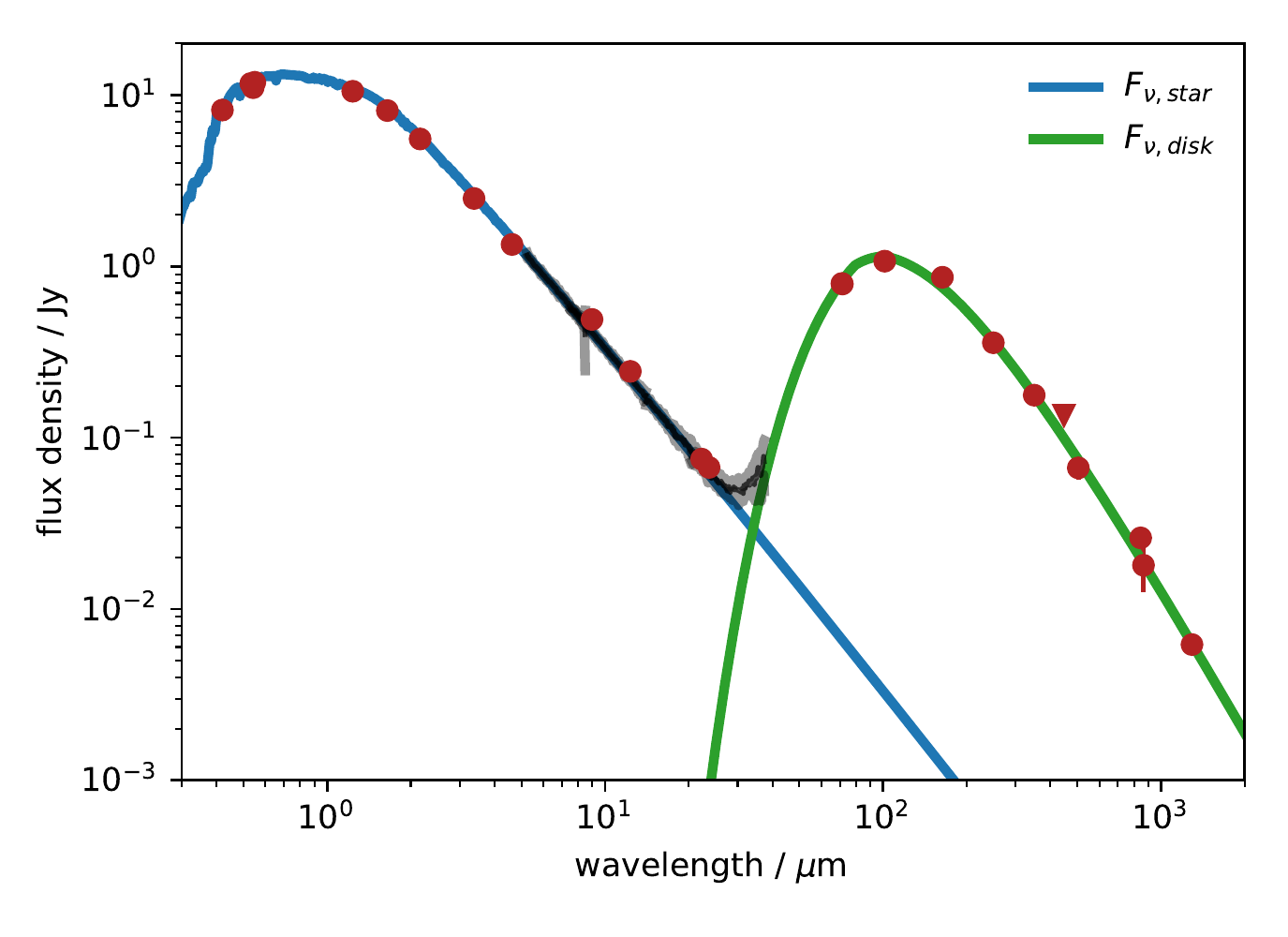}
    \caption{Flux distribution for HD 170773. Dots show photometry and the black and grey lines show the \textit{Spitzer} IRS spectroscopy and uncertainty. The downward pointing triangle shows the JCMT 450 $\mu$m non-detection upper limit. The blue line shows the best-fit stellar photosphere, and the green line the best-fit modified blackbody for the disk.
    }
    \label{fig:SED}
\end{figure}

The 1.3 mm flux density of the disk was estimated in several ways. The flux density of the disk measured in the $>~3\sigma_{\rm RMS}$ region of the ring is 5.04 $\pm$ 0.52 mJy including an absolute flux uncertainty of 10\% added in quadrature. Similarly, the flux density is 5.22 $\pm$ 0.54 mJy when the flux interior to the ring is included in the measurement. The flux density of the $>~2\sigma_{\rm RMS}$ region of the ring together with the interior region is 5.38 $\pm$ 0.56 mJy. This is less than the flux density obtained from visibility modeling (6.2 $\pm$ 0.2 mJy, \S\ref{subsec:modelRes}). The latter extrapolates to baselines shorter than probed by the ACA based on an assumed disk structure to recover emission missing from the image. This measurement (6.2 $\pm$ 0.6 mJy when including the absolute flux uncertainty) is shown, along with other photometry (see Table \ref{tab:photometry} for the complete list) and \textit{Spitzer} IRS spectroscopy of the HD~170773 system \citep{2011ApJSCASSIS}, in Figure \ref{fig:SED}. 

We fit grids of star and disk models to the data, using synthetic photometry of the models to fit photometry, and resampled model spectra to fit the IRS spectrum. The model parameters are the stellar temperature \citep[fit with a PHOENIX model atmosphere,][]{Husser+2013} and normalization (i.e. solid angle), and the disk temperature, normalization, and two “modified blackbody” parameters (the Planck function is divided by $\lambda^\beta$ beyond $\lambda_0\ \mu$m as a simple means to model the inefficient grain emission at long wavelengths). The best fitting model parameters are found with the \texttt{MultiNest} code \citep{Multinest}, with both the stellar and disk parameters found simultaneously.

We find best-ft values of $T_{\rm eff} = 6640$~K for the stellar temperature, and 40~K for the “modified” blackbody (the blackbody function is divided by $\lambda^{0.9}$ beyond 200~$\mu$m) of the disk. The 850 $\mu$m flux from JCMT observations is higher than, but still within $\sim3\sigma$ of the best-fit model. The overall millimeter spectral slope beyond 200 $\mu$m is $ \alpha_{mm} = 2.87 \pm 0.04$. This is comparable to the millimeter spectral slopes for a sample of 15 other disks \citep{macgregor2016}. Here the fractional luminosity of the disk ($L_{disk}/L_{\star}$) is $ f = (5.0 \pm 0.1) \e{-4}$. We note a 3.6$\sigma$ flux excess at 24 $\mu$m that this model does not account for. One possible explanation could be the presence of a warm inner disk which is not detected in our ALMA observations.

\clearpage

\subsection{Disk Modeling and Fitting}\label{subsec:model}

To precisely constrain the fundamental spatial properties of the debris disk, we forward-model the disk surface density profile using an azimuthally symmetric vertically thin Gaussian ring, which is a commonly used model for describing millimeter-resolved debris disks \citep[e.g.][]{macgregoretal2015,boothetal2017,suetal2017,marino2017,marshall2018,Matra+2019BetaPic}. The mean ($\mu$) of the Gaussian represents the radius of the disk at peak surface density, and the standard deviation ($\sigma$) represents the spread of the surface density with respect to the peak. The temperature dependence as a function of radius (assuming blackbody dust grains) is also factored into our surface brightness profile as $S_B \propto B_\nu(T) \propto T(r) \propto \frac{1}{\sqrt{r}}$, where $B_\nu(T)$ is the Planck function which, for long wavelengths, is proportional to the inverse square root of the radius. The complete radial dependence of surface brightness in our model is given by:
\begin{equation}\label{eqn:model}
S_B (f_0,\sigma,\mu; r) \propto \frac{f_0}{\sqrt {r}}e^{{{-({r - \mu })^2 }/{2\sigma ^2}}}
\end{equation}

This model is parameterized by $\sigma$, $\mu$, and $f_0$, where $f_0$ represents the integrated flux of the belt. In addition to these model parameters, we also account for four additional parameters which describe the line of sight inclination ($i$), position angle of the major axis measured east of north ($PA$), and offsets of the belt's geometric center from the phase center of the observation ($\Delta$$RA$ and $\Delta$$Dec$). We have accounted for potential systematic errors in the values of the visibility weights delivered by ALMA, which have been found in other datasets \citep[e.g.][]{kennedy2018,marino2018}, and which could also otherwise mean that our uncertainties are underestimated. This is done by including a free parameter multiplied by the weights of each of the three observational datasets.

The \texttt{galario} \citep{GALARIO} package is utilized to Fourier transform the model image at the u-v locations of our ALMA data to  calculate the ${\chi}^2$ of the model given the data. To derive the best-fitting synthetic model visibility dataset, the posterior probability distributions of our model parameters are explored using the \texttt{emcee} \citep{foremanmackey13} package, the affine-invariant ensemble sampler implementation of Markov Chain Monte Carlo \citep{goodmanWeare2010} in Python. We use a likelihood function proportional to $\exp(-\chi^2/2)$ and use linearly uniform priors on all our model parameters $f_0$, $\sigma$, $\mu$, $i$, $PA$, $\Delta$$RA$ and $\Delta$$Dec$. We initialize an ensemble of 1024 walkers each sampling from the parameter space for 1.25$\e4$ time steps with a burn-in strip size of 2.5$\e3$. We assessed the convergence of the Markov chains by comparing their length to their integrated autocorrelation times for each parameter (estimated with the \texttt{emcee} package). We find that all the chains are at least $\sim$67 times their integrated autocorrelation time, ensuring convergence. The chains were also visually inspected to confirm that a steady state was reached.

\begin{deluxetable*}{cccc}
\tablehead{
\colhead{Parameter} & \colhead{Short Description} & \colhead{Best-Fit} & \colhead{68\% Confidence Interval} 
}
\caption{Fundamental disk parameter constraints. We report the best-fit values as the median value of the resulting posterior distribution, and the 68\% interval as the confidence interval. The parameters directly involved in the \texttt{emcee} runs lie above the solid line. Below the solid line are the parameters derived from the \texttt{emcee} results, where the radius (R) is $\mu$ converted to au and the disk width $\Delta$R is the FWHM ($2\sqrt{2\ln 2}\sigma$) converted to au.\label{tab:res}} 
\startdata
$f_0$ [mJy]& disk flux &6.2&(+0.2,-0.2)\\
$\sigma$ [\arcsec]&standard deviation of radial surface density distribution&0.82&(+0.04,-0.04)\\
$\mu [\arcsec]$&disk radius at peak surface density&5.20&(+0.05,-0.08)\\
$i$ [$^{\circ}$]&line-of-sight inclination&33&(+1,-2)\\
$PA$ [$^{\circ}$]&position angle (east of north)&114&(+2,-3)\\
$\Delta$$RA$ [\arcsec]&right ascension offset&-0.06&(+0.05,-0.05)\\
$\Delta$$Dec$ [\arcsec]&declination offset&0.03&(+0.04,-0.04)\\
\hline
R [au]&disk radius at peak surface density&193&(+2,-3)\\ 
$\Delta$R [au]&width of disk&71&(+4,-4)\\
\enddata
\end{deluxetable*} 
\begin{figure*}
  \includegraphics[width=7in]{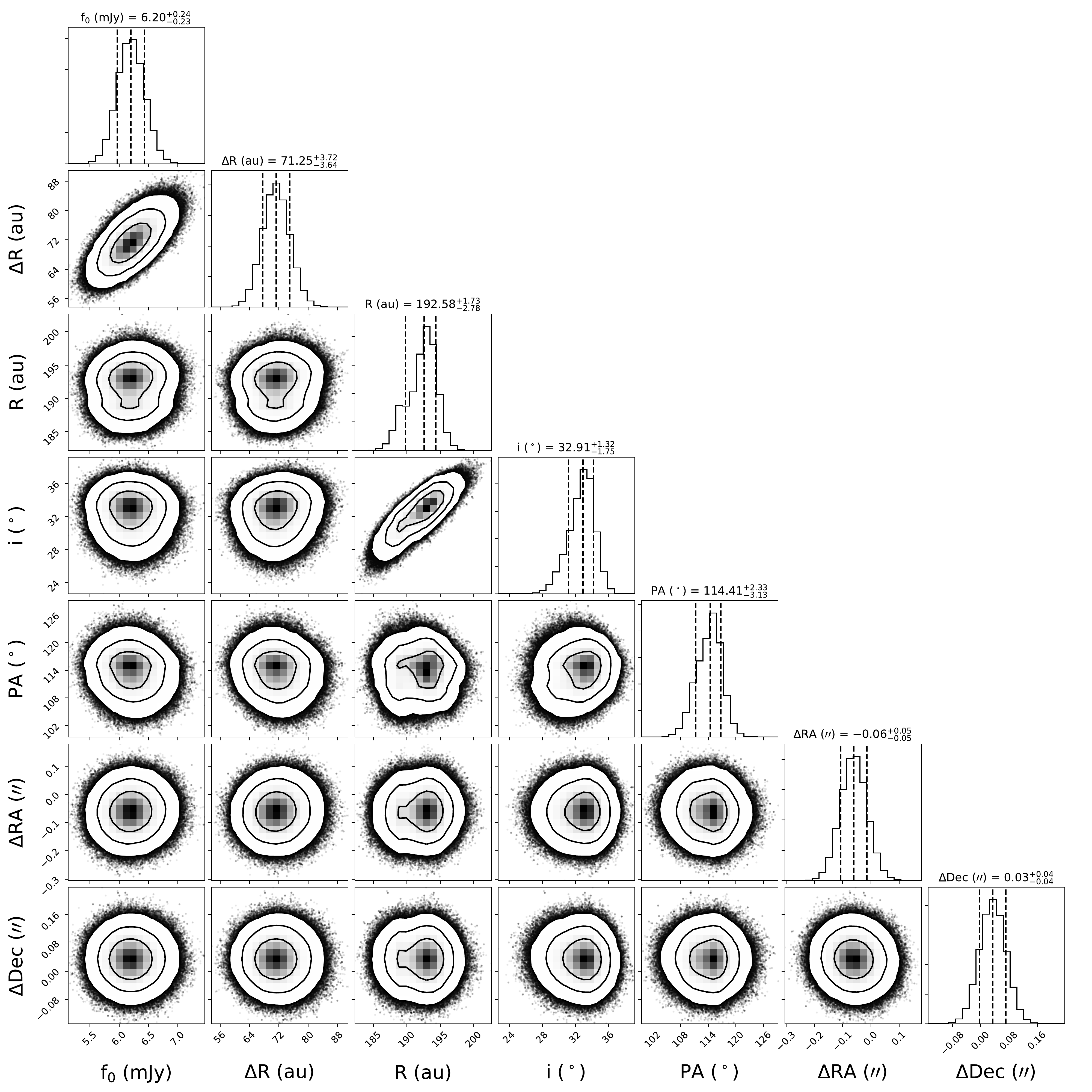}
  \centering
  \caption{Cornerplot summary of the \texttt{emcee} run results showing the marginalized posterior probabilities for the disk parameters (1-D histograms) as well as the two-dimensional projections for each combination of the disk parameters (2-D histograms). The 1-D posterior probabilities have their 16\%, 50\%, and 84\% quantiles displayed as vertical dashed lines and with their values listed above each respective distribution. The contours of the 2-D histograms are displayed for 68\%, 95\%, and 99.7\% density levels. 
    }
    \label{fig:corner}
\end{figure*}

\begin{figure*}
    \includegraphics[trim=0cm 5.5cm 0cm 5.5cm, clip, width=7.1in]{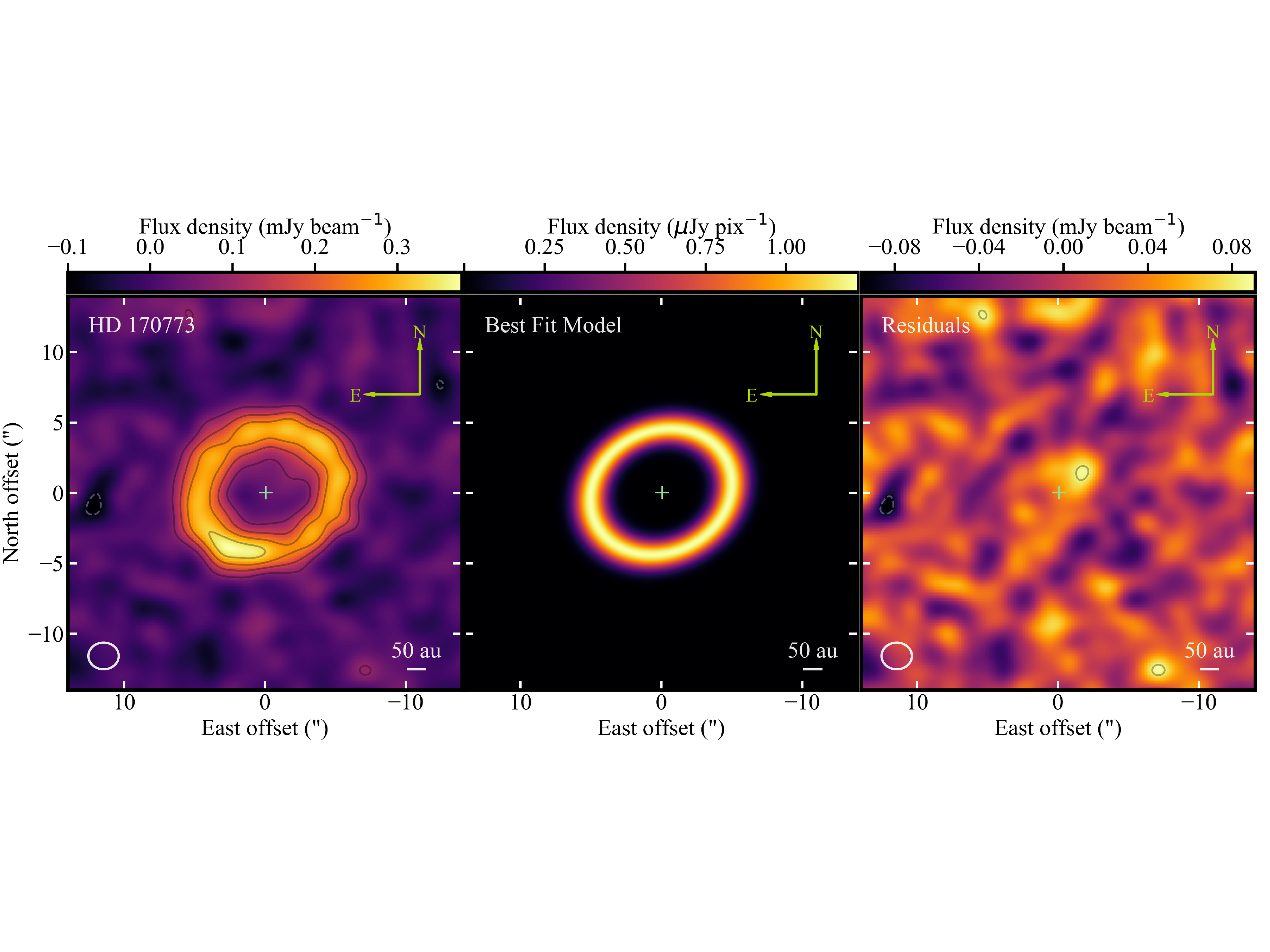} 
    \centering
    \caption{Synthetic, full-resolution model of the HD 170773 debris disk (center) created using the best-fit parameters from the \texttt{emcee} run. Residual image of the disk (right) obtained after subtracting the best-fit model visibilities from observed visibilities and inverse transforming into image space. Contours for the residual image are drawn at the levels: $[-3,3] \times \sigma_{\rm RMS}$. The continuum image (left) is the same as Figure \ref{fig:emission}. The plus sign indicates the best-fit geometric center of disk.
    }
    \label{fig:res}
\end{figure*}
\subsection{Model Fitting Results}\label{subsec:modelRes}
Figure \ref{fig:corner} shows a corner plot summary of the \texttt{emcee} results. We summarize the best-fit values of the disk parameters (taken as the median value for each respective posterior distribution) together with their 68\% confidence intervals in Table \ref{tab:res}. Our constraint for $\mu$ yields a radius at peak emission constraint of 193$^{+2}_{-3}$ au. The difference of this measurement compared to the constraints of \citet{mooretal2015} and \citep{holland2017} is likely due to differences in modeling formalism (e.g., \citet{mooretal2015} use an annulus of constant surface brightness), although the radii are still broadly consistent. The inclination and position angle of 33$^{\circ}$$^{+1}_{-2}$ and 114$^{\circ}$$^{+2}_{-3}$ are consistent with the constraints of \citet{mooretal2015}, who reported the parameters as 31.3$^{\circ}$$\pm$1.7 and 118.3$^{\circ}$$\pm$3.2, respectively. The phase center offsets ($\Delta$$RA$ and $\Delta$$Dec$) are consistent with zero, indicating the disk's geometrical center is consistent with the stellar location, in contrast with some other debris disks found to be eccentric \citep[e.g. Fomalhaut,][]{Kalas2005,macgregor2017}. The width of the disk, interpreted as the FWHM of the Gaussian surface density distribution, is 71 $\pm$ 4 au. This makes the disk somewhat narrow ($\Delta R/R = 0.37$), which is common given the current observations of debris disks \citep{hughes18}.

\begin{figure}
    \includegraphics[width=3.5in]{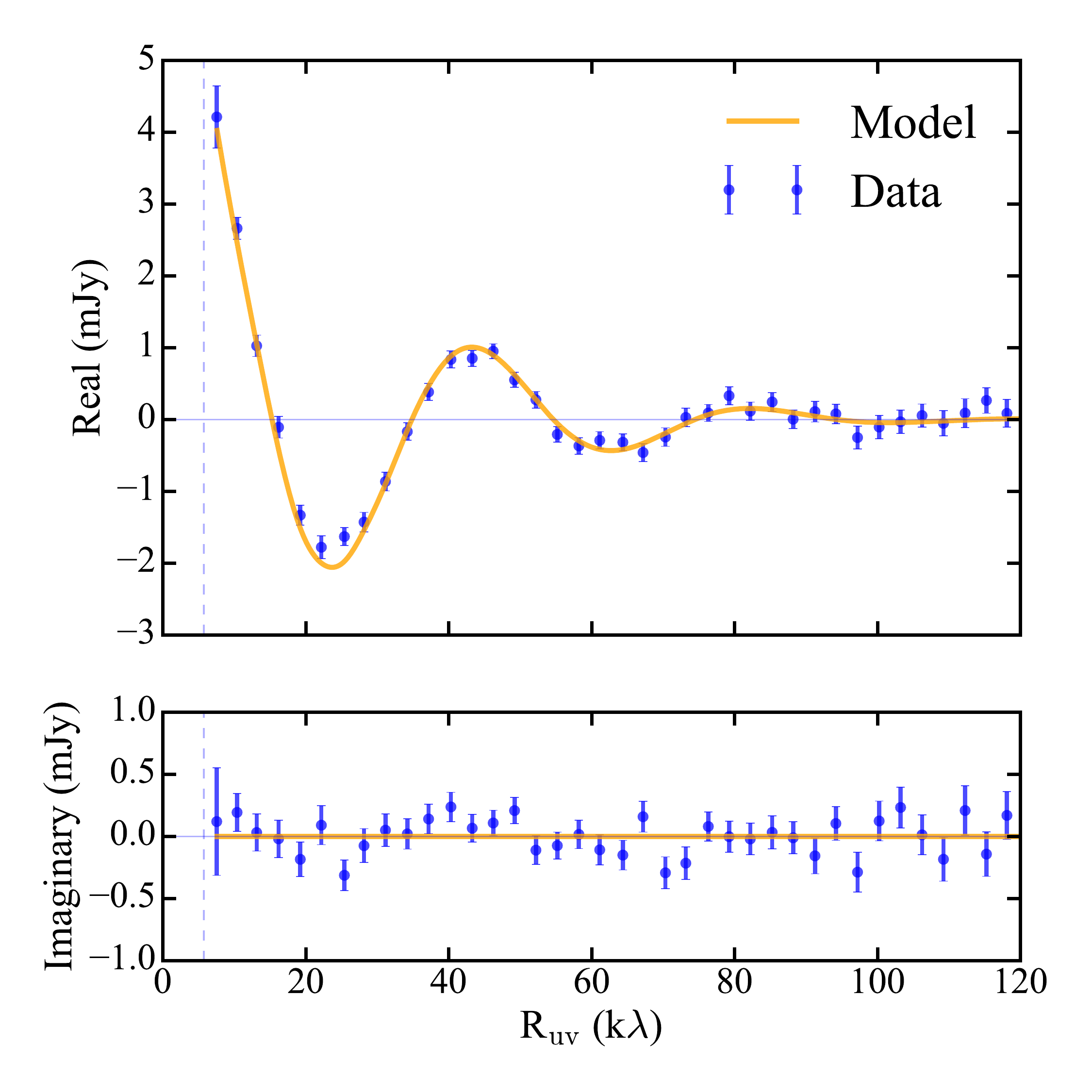}
    \hspace{+0.5cm}
    \caption{Real and imaginary part of the interferometric visibilities as a function of u-v distance deprojected assuming the belt’s best-fit position angle of 114$^{\circ}$ and inclination of 33$^{\circ}$. The best-fit model (orange line) is consistent with the data (blue error bars) within the uncertainties. The imaginary part of the visibilities being consistent with zero supports the lack of significant evidence for disk asymmetries.
    }
    \label{fig:deprojvis}
\end{figure}

In Figure \ref{fig:res} we show the image of the synthetic best-fit model of the disk along with the corresponding image of the residual visibilities after subtraction of this best fit model. The residual image is consisent with noise and no significant emission at the $>4\sigma_{\rm RMS}$ level is present at the disk location, indicating that the best-fit model parameters are consistent with the data and also confirming the lack of significant evidence for the possible asymmetry noted in \S\ref{subsec:cont}. We only note a marginal positive 3$\sigma_{\rm RMS}$ residual to the NW and slightly offset from the stellar location. We also show the deprojected visibility plot of the data and best-fit model as Figure \ref{fig:deprojvis}. The imaginary visibilities being consistent with zero further supports the axisymmetric model. 

The asymmetric shape of the radius (as well as $i$ and $PA$ ) posterior distributions in Figure \ref{fig:corner}, in addition to a potential marginal over subtraction of the ring in Figure \ref{fig:res} (right panel), suggests a potential skewness of the Gaussian distribution we used as our parametric model for the surface density. We addressed this by remodeling the disk with more freedom for the standard deviation of the Gaussian profile to differ in the inward and outward radial directions. Here $\sigma$ is replaced by two free parameters where now $\sigma_{in}$ is defined for $r<r_0$ and $\sigma_{out} $ for $r>r_0$. This second model is otherwise identical to the first. While visibility fitting with this asymmetric surface density model resulted in a marginally better ${\chi}^2$ value, we found that due to the additional free parameter it did not describe the data significantly better than the symmetric density model. This was assessed using the Bayesian Information Criterion \citep[BIC,][]{BICCitation}, where we find a $\Delta BIC$ value of 7.3 favoring the symmetric density model.

\subsection{Constraints on CO and CN Line Emission}
No clear detection is seen in the datacubes around the frequency of the CO J=2-1 line and the CN N=2-1, J=5/2-3/2 transition, where the latter is composed of three, blended, hyperfine components (F = 7/2-5/2, 5/2-3/2 and 3/2-1/2). Given the disk is resolved over many spatial, and potentially spectral, resolution elements, we employ the spectro-spatial filtering technique of \citet{matra2015,matra2017b} to boost the SNR by assuming CO and CN are co-located with the dust and in Keplerian velocity (where both rotation directions were tested) around the star (of mass 1.29 M$_{\odot}$, \S\ref{subsec:stellarParam}). This is what is expected for gas released from exocometary ices within the collisional cascade that also produces the dust, as observed in several other systems \citep[e.g.][]{marino2016,matra2017b}.

No detection is achieved; we set an upper limit (3$\sigma$) of 35 and 47 mJy km/s on the integrated line flux of the CO and CN transitions, respectively. This was calculated from the RMS of the spectro-spatially filtered spectra, multiplied by the effective bandwidth of the instrument (2.667 times the channel width) assuming, as expected, that the spectro-spatially filtered line is close to unresolved spectrally. This uncertainty was added in quadrature to a 10\% absolute flux uncertainty expected from ALMA observations.

We then derived CO and CN gas mass upper limits from the observed fluxes, using the non-local thermodynamic equilibrium (NLTE) excitation code of \citet{matra2015,matra2018a} in the optically thin assumption. We explore the full range of collider densities between the regime where excitation is dominated by collisions (LTE) and radiative absorption/emission, and temperatures between 10 and 250 K. This allows us to derive upper limits of 1-14$\times10^{-6}$ and 1.3-8.0$\times10^{-8}$ M$_{\oplus}$ on the CO and CN gas masses, respectively.

We then assume that any gas that may be present is being released from exocometary ice through a steady state collisional cascade and destroyed through photodissociation at the same rate as it is produced. Photodissociation at the 193 au radius of the belt around an F star such as HD~170773 is dominated by the interstellar radiation field (ISRF), leading to photodissociation timescales of $\sim$120 and $\sim61$ years for CO and CN \citep{heays2017}. Given that HCN is the main parent molecule producing CN via photodissociation, we can use CN to probe the exocometary HCN ice content \citep{matra2018a}.

As long as all CO and HCN are released from solids by the time these are ground down to the smallest size in the collisional cascade, the CO and/or HCN release is proportional to the mass loss rate of the belt \citep[see \S\ref{sec:collcasc} in this work and Eq. 2 in ][]{matra2017b}, and can be used to extract the ice mass fraction in exocomets. Around HD 170773, we estimate an upper limit to the CO and HCN exocometary mass fraction of $<77$\% and $<3$\%, respectively. The CO limit is consistent with CO mass fractions of a few to a few tens of percent derived from detection around other exocometary belts, as well as Solar System comets \citep[e.g.][]{mumma&charnley2011} assuming a rock/ice ratio of $\sim$4 as measured in comet 67P \citep{rotundi2015}.

\begin{deluxetable}{ccc}[b]
\tablehead{
\colhead{Stellar Parameter} & \colhead{Derived Value} & \colhead{68\% Confidence Interval}
}
\caption{Stellar parameters and 68\% uncertainties.\label{tab:stellarprop}} 
\startdata
Age [Gyr]&1.5&(+1.2,-0.7)\\
Mass [$\Msun$]&1.29&(+0.08,-0.08)\\
log(L) [$\Lsun$]&0.558&(+0.006,-0.006)\\
$\rm T_\mathrm{eff}$ [K]&6551&(+32,-32)\\
logg [cgs]&4.21&(+0.04,-0.04)\\
Radius [$\Rsun$]&1.477&(+0.022,-0.022)\\
\enddata
\end{deluxetable} 
\subsection{Stellar Parameters}\label{subsec:stellarParam}
The stellar parameters of HD 170773, i.e., age, mass, luminosity, effective temperature, surface gravity, and radius, were inferred by employing the absolute G magnitude (6.1040$\pm$0.0004 mag, obtained from the apparent G magnitude and the parallax) and BP$-$RP color from Gaia DR2 \citep[0.5691$\pm$0.0059 mag,][]{gaia,gaiaDR2_2018,Lindegren2016}, as well as [Fe/H] (assumed to be solar, 0.0 $\pm$ 0.20 dex), all three as input parameters using the Bayesian approach applied in \citet{delBurgo&AllendePrieto2016,delBurgo&AllendePrieto2018}. We inferred that HD 170773 is most likely a main sequence star and show the derived parameters in Table \ref{tab:stellarprop}.

In order to infer the stellar parameters, we downloaded and arranged a grid of \texttt{PARSEC} isochrones \citep[version 1.2S,][]{Bressan2012,chen2014,chen2015,Tang2014}, using the synthetic photometry from \citet{Evans+2018}. The iron-to-hydrogen ratio [Fe/H] ranges from -2.18 to 0.50, in steps of 0.02 dex, the age goes from 200 Myr to 13.5 Gyr, in steps of 5\%, and the initial mass ranges from 0.09 $\Msun$ to the highest mass established by the stellar lifetimes, in irregular steps that properly sample the slow and fast evolutionary phases. The absolute maxima for the initial mass and actual mass in the grid are 350.0 $\Msun$ and 345.2 $\Msun$, respectively. For a more detailed description, see \citet{delBurgo&AllendePrieto2018}.

\section{Discussion}\label{sec:discuss}

\subsection{Collisional Cascade Status}\label{sec:collcasc}
The blowout of the smallest dust grains in the debris disk due to radiation pressure, which are themselves created by the destruction of larger bodies, results in mass loss over time. Consequently, older debris disks will be less luminous and are harder to detect than younger disks. \citet{matra2017b} derive a simple equation for the mass loss rate of these smallest grains assuming a steady-state collisional cascade model, given as $\dot M_{D_{min}} = 1.2\e3R^{1.5} \Delta R^{-1} f^{2}L_{\star} {M_{\star}}^{-0.5}$ where R is in au, $L_{\star}$ is in \Lsun, and $M_{\star}$ is in \Msun.
Adopting the best-fit values of these parameters from \S\ref{sec:res}
yields a mass loss rate of $\dot M_{D_{min}} = 3.6\e{-2}\ \rm M_{\oplus}$Myr$^{-1}$  for HD~170773. Assuming that mass loss has been ongoing at this constant rate for the age of the star, the total mass lost for HD 170773 is $54\ \rm M_{\oplus}$ when using an age of 1.5 Gyr (\S\ref{subsec:stellarParam}). We note that there is substantial uncertainty in the age of HD 170773, as pinning down precise stellar ages is generally difficult. If the age of HD 170773 is as young as 200 Myr \citep{zuckerman+song2004}, then the total mass lost is 7.2 $\rm M_{\oplus}$.

Assuming a millimeter dust opacity of $\kappa _{\nu} = 2.3\ \rm cm^2 g^{-1}$ yields the mass in millimeter dust grains as $7.4\e{-1}\ \rm M_{\oplus}$ \citep[Eq. 7 in][]{wyatt2008rev}. However, in order to produce the dust we see today, the size distribution must extend to much larger bodies, which will make the total belt mass much higher. For a steady-state size distribution described by a power law where $n(D) \propto D^{-3.5}$, the total mass of the collisional cascade can be linked to the size $D_c$ km of the largest bodies feeding the cascade \citep[knowing the fractional luminosity, radius and minimum grain blowout size, see Eq. 15,][]{wyatt2008rev}. This leads to the expression $ M_{tot} = 75 \sqrt{D_c}\ \rm M_{\oplus}$ for HD~170773. In addition, we can calculate the collisional timescale $t_c$ Myr of these largest bodies of size $D_c$ for a total solid mass within the cascade $M_{tot}$ \citep[as a function of the known, or assumed, spatial properties of the disk, host star mass, planetesimal strength, and mean planetesimal eccentricity, see Eq. 16,][]{wyatt2008rev}. Assuming a planetesimal strength of 150 J/kg and a mean planetesimal eccentricity of 0.05 yields $ M_{tot} = 2.8\e{4} \times t_c^{-1} D_c\ \rm M_{\oplus}$ for HD 170773. 

Combining the two equations for $M_{tot}$ leads to $D_c = 7\e{-6} \times t_c^2$ km. Since the largest bodies of size $D_c$ participating in the collisional cascade will be those whose timescale $t_c = t_{age}$ (assuming the collisional cascade has been ongoing for the age of the star), we let $t_c = 1.5$ Gyr to find the size of the largest bodies needed to produce the dust we observe at the star's age. Here that size is $D_c = 16$ km, which we in turn use to find a total mass of $ M_{tot} = 299$ $\rm M_{\oplus}$. This estimate should be considered a lower limit on the total disk mass since there could be larger planetesimals in the belt that are yet to suffer a collision and hence are not participating in the cascade. For an age range of 0.8-2.7 Gyr (Table \ref{tab:stellarprop}), the size of the largest bodies in the collisional cascade is 5-52 km, which implies total disk masses of at least 175-539 $\rm M_{\oplus}$. 

To check whether such a large belt mass is reasonable, we compare it with the expectation from the Minimum Mass Solar Nebula \citep[MMSN,][]{Weidenschilling1977,Hayashi1981}, linearly rescaled to account for a stellar host of mass 1.29 $\rm M_{\oplus}$ \citep[as done in, e.g.][]{Kenyon&Bromley2008}.
At 193 au around HD 170773, the expected MMSN-like surface density of solids would be $1.4\times10^{-2}$ g cm$^{-2}$ \citep[Eq. 2.5 in][]{Hayashi1981}, which leads to an estimated MMSN-like disk mass of 46 $\rm M_{\oplus}$ for a simple belt of large planetesimals of width 71 au and MMSN-like surface density. We therefore conclude that the large dust content and radius of the HD 170773 disk requires feeding from a planetesimal belt at least a few to an order of magnitude more massive than expected from a MMSN-like protoplanetary disk.

While keeping in mind the significant uncertainties in the inputs of the mass calculation (dominated by the system age, and by the unknown planetesimal strengths and eccentricities), the high masses obtained for the HD 170773 belt reinforce a disk mass problem recently highlighted for bright debris disks \citep{Krivov2018,kennedy2018}. To reconcile observed belt masses with the expectation from the MMSN and observed protoplanetary disks, belts could either (1) have lower dynamical excitation than currently assumed (leading to lower eccentricities and higher dust masses at later ages); (2) have different strengths and size distributions than typically assumed; (3) have been collisionally evolving for a time shorter than the system age \citep[requiring delayed stirring, e.g.][]{Kenyon&Bromley2008}; or (4) have sources of additional dust production beyond catastrophic collisions within the cascade considered here.

\subsection{Radius-Luminosity Relationship}
\begin{figure}[b]
    \includegraphics[trim=0.75cm 1cm 0cm 0.65cm, clip, width=3.5in]{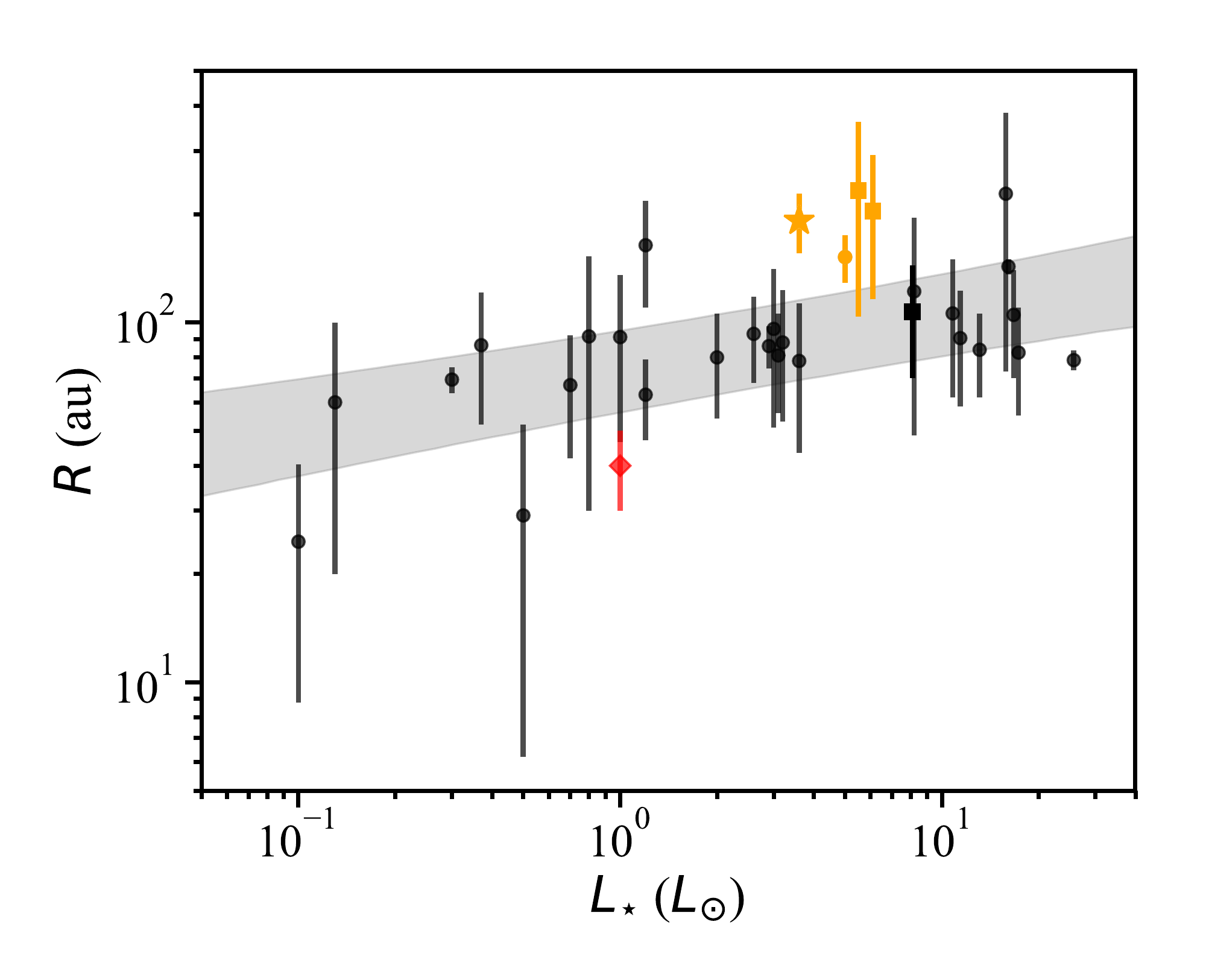}
    \caption{Radius-Luminosity relation that expands upon the one published in \citet{matra2018b}. The black points represent the published millimeter-resolved belt radii and corresponding belt widths, compared to the host star luminosity. The grey slope represents the effective 1$\sigma$ confidence interval for a range of power laws which describe the correlation. The four potential outliers are identified in orange, where HD 170773 is denoted with a star symbol. Systems that host directly-imaged giant planets are denoted with a square. The red diamond represents our Solar System's Kuiper Belt. 
    }
    \label{fig:RLPlot}
\end{figure}
\citet{matra2018b} analyzed the 26 published debris disks resolved at millimeter wavelengths and found a significant correlation between their radii and host star luminosities. We update the $R-L_{\star}$ relation by incorporating HD 170773 and any new studies of millimeter-resolved debris disks published since that time. The new additions are HD 32297 \citep{MacGregor+2018}, HR 4796A \citep{kennedy2018}, HD 92945 \citep{Marino+2019}, TWA 7 \citep{Bayo+2018,Matra+2019TWA7}. The disks from the original sample with updated spatial properties are HD 107146 \citep{marino2018}, HD 61005 \citep{MacGregor+2018}, $\beta$ Pic \citep{Matra+2019BetaPic}, HD 131835 \citep{Kral+2018} and HR 8799 \citep{Wilner18}. We find the updated correlation parameters to be $R_{1L_{\odot}
}=74 \pm 7$ au, $\alpha=0.16 \pm 0.05$, and $f_{\Delta R} = 0.25^{+0.08}_{-0.06}$
\citep[for more details on fitting the $R-L_{\star}$ relation, see \S 2 in][]{matra2018b}. 

The HD 170773 debris disk is larger in radius than typical disks around late F/early G type stars located between 2-4 $\Lsun$ on the updated $R-L_{\star}$ relation plot (Figure \ref{fig:RLPlot}). The disk instead lies closer to three other potentially outlying early F/late A disks, which are HR 8799 \citep{booth2016,Wilner18}, HD 95086 \citep{suetal2017}, and $\eta$ Crv \citep{marino2017}. The significance of each of the four disks HD~170773, HR~8799, HD~95086 and $\eta$~Crv being outliers are 3.5$\sigma$, 3.9$\sigma$, 3.4$\sigma$, and 2.2$\sigma$, respectively. This was evaluated by comparing the best-fit disk radius to the probability distribution of radii at the corresponding host star luminosity given the updated correlation parameters. While this indicates that these belts may truly form an outlying group, we caution that this conclusion is sensitive to the intrinsic scatter of radii about the best-fit $R-L_{\star}$ model being modelled as a Gaussian (as opposed to, e.g., a top-hat) distribution \citep[\S 2,][]{matra2018b}. 

An interesting property of this potentially outlying group is that half of the members, namely HR 8799 \citep{marios2008,marois2010} and HD 95086 \citep{rameau2013}, harbor giant planets that have been directly imaged. The two systems with detected giant planets have young age constraints ($\sim$40 Myr for HR 8799 \citep{zuckerman2011} and $\sim$17 Myr for HD 95086 \citep{Meshkat+2013}) while the two with non-detections have older age constraints ($\sim$1.3 Gyr for $\eta$ Crv \citep{Mallik+2003} and $\sim$1.5 Gyr for HD 170773 (\S \ref{subsec:stellarParam})), suggesting that the non-detections may be due to the giant planets having cooled and become too faint for direct detection \citep[e.g., for $\eta$ Crv see][]{Lafreniere+2007}. By contrast, only two of the 27 $R-L_{\star}$ members not in this group host imaged planets (barring the Solar System's Kuiper Belt for this consideration). These members are Fomalhaut \citep{Kalas+2008} and $\beta$ Pic \citep{lagrangeetal2009,Dupuy+2019}, though the imaged companion of Fomalhaut is not likely in the giant planet mass regime \citep[e.g.][]{Kalas+2013,Beust+2014,Lawler+2015}. This tentative distinction could be alluding to some relationship between disk radii and long-period giant planet frequency, but more millimeter observations of disks and complementary direct imaging exoplanet surveys are needed to make a more robust approach to the problem.

\subsection{Exploring Hypothetical Exoplanetary System Architecture}

Stars that host both a debris disk and one or more exoplanets serve as critical test beds for studying the formation and evolution of planetary systems. Giant planets are present around at least $\sim6\%$ of stars with a detected debris disk compared to at least $\sim 0.7\%$ of stars without a detected debris disk \citep{meshkatetal2017}. To determine whether this distinction is due to observational bias or due to some intrinsic physical relation requires more observational efforts to detect disks and exoplanets for a larger sample of stars.

A direct imaging survey carried out with Gemini Observatory (NICI Campaign) resulted in no detection of giant planets for HD 170773, indicating that any present companions were below the detection limits of 9.0 \Mjup\ at 148 au and 13.4 \Mjup\ at 74 au \citep[][hot-start models are used]{wahhaj2013}. The age used to calculate these hot-start mass upper limits was 200 Myr, which is less than the age we derive in this study and could thus mean that the mass upper limits may be higher. If an undetected planet around HD 170773 is both massive enough and orbiting at the necessary proximity to the inner edge of the disk to be clearing disk mass in its chaotic zone, then the inner radius of the disk, planet semi-major axis, and planet mass can be related by $R_{in} = a_{pl} + 5a_{pl}(M_{pl}/3M_{\star})^{1/3}$ \citep[][assuming a non-eccentric orbit]{pearceandwyatt2014}. Assuming this scenario and comparing against a linear interpolation of the published planet mass upper limits yields an estimate of 91 au for the minimum planet semi-major axis. While a planet can approach an infinitesimally small mass at decreasingly short separations between it and the disk and still be consistent with these conditions, if we consider only the giant planet mass regime then a 1 \Mjup\ planet could plausibly be orbiting at 120 au. Future, deeper imaging with higher contrast sensitivity will be needed to reveal the presence of such an orbiting companion.

\section{Conclusion}
We used ALMA to obtain the first millimeter-resolved observations of the dust and gas around HD 170773 as part of the REASONS survey. We forward-modeled the disk as an axisymmetric thin Gaussian ring and found the disk width to be 71$^{+4}_{-4}$ au and the radius from the host star to be 193$^{+2}_{-3}$ au. The spatial properties are consistent with previous studies of this disk, and reveal that HD 170773 hosts a large and narrow debris disk when compared to the currently known disk population.

We also searched for any CO and CN gas released by exocomets in the system. We set upper limits on their gas mass, which allow us to constrain the mass fraction of CO and HCN ice in exocomets to $<77$\% and $<3$\%, respectively. These upper limits still allow for HD~170773 to be hosting icy exocomets with compositions analogous to the Solar System and other known gas-bearing exocometary belts.

The disk characteristics were used together with constraints on the stellar parameters from \textit{Gaia} DR2 to estimate some of the system properties. We found that bodies must be at least 16 km in diameter to sustain a steady-state collisional cascade producing the currently observed dust levels after 1.5 Gyr of evolution. This 16 km size would lead to a total mass in the collisional cascade of $\sim300$ $\rm M_{\oplus}$. This is almost an order of magnitude larger than expected from a MMSN-like protoplanetary disk, which (barring significant uncertainties in some of the assumed parameters) provides further support for the presence of a disk mass problem for bright debris disks \citep{Krivov2018, kennedy2018}. 

In the context of the $R-L_{\star}$ relation of planetesimal belts from mm-wave imaging, HD 170773 is part of a group of potentially outlying large F star disks. Interestingly, two of the four potential outliers in this group also host directly imaged long-period giant planets (versus 1/27 in the remaining belt population), which may suggest a relationship between the frequency of long-period giant planets and the presence of large debris disks. Around HD 170773, we find that a hypothetical long-period giant planet clearing material inwards of the disk's inner edge should lie beyond 91 au to remain below current direct imaging detection limits.

REASONS and other future surveys will play a vital role in characterizing the spatial properties of debris disks. These surveys should be complemented by direct imaging surveys to find systems which host both exoplanets and a debris disk, providing an invaluable laboratory to further analyze the dynamics of exoplanetary systems.

\vspace{5mm}
\acknowledgments
\textit{Acknowledgments}. AGS acknowledges support from the SAO REU program, funded in part by the National Science Foundation REU and Department of Defense ASSURE programs under NSF Grant no.\ AST-1659473, and by the Smithsonian Institution. LM acknowledges support from the Smithsonian Institution as a Submillimeter Array (SMA) Fellow. GMK is supported by the Royal Society as a Royal Society University Research Fellow. CdB acknowledges the funding of his sabbatical position through the Mexican national council for science and technology (CONACYT grant CVU No.448248). MB acknowledges support from the Deutsche Forschungsgemeinschaft through project Kr 2164/15-1. JMC acknowledges support from the National Aeronautics and Space Administration under Grant No. 15XRP15\_20140 issued through the Exoplanets Research Program. CLD acknowledges support from the ERC Starting Grant “ImagePlanetFormDiscs” (grant agreement No. 639889). JPM acknowledges research support by the Ministry of Science and Technology of Taiwan under grants MOST104-2628-M-001-004-MY3 and MOST107-2119-M-001-031-MY3, and Academia Sinica under grant AS-IA-106-M03.

This paper makes use of the following ALMA data: JAO.ALMA\#2018.1.00200.S. ALMA is a partnership of ESO (representing its member states), NSF (USA) and NINS (Japan), together with NRC (Canada), MOST and ASIAA (Taiwan), and KASI (Republic of Korea), in cooperation with the Republic of Chile. The Joint ALMA Observatory is operated by ESO, AUI/NRAO and NAOJ. The National Radio Astronomy Observatory is a facility of the National Science Foundation operated under cooperative agreement by Associated Universities, Inc. 

This work has made use of data from the European Space Agency (ESA) mission {\it Gaia} (\url{https://www.cosmos.esa.int/gaia}), processed by the {\it Gaia}
Data Processing and Analysis Consortium (DPAC, \url{https://www.cosmos.esa.int/web/gaia/dpac/consortium}). Funding for the DPAC has been provided by national institutions, in particular the institutions participating in the {\it Gaia} Multilateral Agreement. 

This research has made use of the NASA Exoplanet Archive, which is operated by the California Institute of Technology, under contract with the National Aeronautics and Space Administration under the Exoplanet Exploration Program.

This research has made use of the SIMBAD database, operated at CDS, Strasbourg, France. 

This research has made use of NASA's Astrophysics Data System Bibliographic Services.
\vspace{5mm}
\facility{ALMA}
\software{\texttt{CASA} \citep{CASA} , \texttt{galario} \citep{GALARIO}, \texttt{emcee} \citep{foremanmackey13}, \texttt{matplotlib} \citep{matplotlib}, \texttt{corner} \citep{corner}, \texttt{astropy} \citep{astropy,astropy2018}, \texttt{numpy} \citep{NumPy},  \texttt{scipy} \citep{scipy}, \texttt{multinest} \citep{Multinest}}
\bibliographystyle{aasjournal}
\bibliography{ms}
\end{document}